\documentclass[aps,pra,twocolumn,superscriptaddress,longbibliography]{revtex4-2}
\usepackage{standalone}
\usepackage[utf8]{inputenc}
\usepackage[english]{babel}
\usepackage{amssymb,amsthm,amsmath,amstext,amsbsy,amsopn,mathrsfs}
\usepackage{bbm}
\usepackage{bm}
\usepackage{nicefrac}
\usepackage{suffix}
\usepackage{array}
\usepackage{slashed}
\usepackage{leftidx,comment}
\usepackage{graphicx}
\usepackage[percent]{overpic} 
\usepackage{environ}
\usepackage{mathtools}
\usepackage{xspace}
\usepackage{xcolor}
\usepackage{scalerel,stackengine}
\usepackage{physics}
\usepackage{braket}
\usepackage{isotope}
\usepackage{float}
\usepackage[OT1]{fontenc}
\usepackage{enumitem}
\usepackage{soul}
\usepackage{multirow, makecell}
\usepackage[driverfallback=dvipdfm]{hyperref}
\usepackage{upgreek}
\hypersetup{colorlinks=true,breaklinks,urlcolor=blue,linkcolor=blue,citecolor=blue}

\begin{document}

\title{The Three-Body Limit Cycle: Universal Form for General Regulators}

\author{Langxuan Chen}
\thanks{These authors contributed equally to this work.}
\affiliation{Department of Physics, Fudan University, Shanghai, 200438, China}

\author{Feng Wu}
\thanks{These authors contributed equally to this work.}
\affiliation{Université Paris-Saclay, CNRS/IN2P3, IJCLab, 91405 Orsay, France}

\author{Xincheng Lin}
\affiliation{Department of Physics and Astronomy, North Carolina State University,
Raleigh, North Carolina 27695, USA}

\author{Sebastian König}
\affiliation{Department of Physics and Astronomy, North Carolina State University,
Raleigh, North Carolina 27695, USA}

\author{Ubirajara van Kolck}
\affiliation{European Centre for Theoretical Studies in Nuclear Physics and
Related Areas (ECT*), Fondazione Bruno Kessler, 38123 Villazzano (TN), Italy}
\affiliation{Université Paris-Saclay, CNRS/IN2P3, IJCLab, 91405 Orsay, France}
\affiliation{Department of Physics, University of Arizona, Tucson, AZ 85721, USA}

\author{Pengfei Zhang}
\email{PengfeiZhang.physics@gmail.com}
\affiliation{Department of Physics, Fudan University, Shanghai, 200438, China}
\affiliation{State Key Laboratory of Surface Physics, Fudan University, Shanghai, 200438, China}
\affiliation{Hefei National Laboratory, Hefei 230088, China}

\begin{abstract}
    The Efimov effect, a remarkable realization of discrete scale invariance, emerges in the three-body problem with short-range interactions and is understood as a renormalization group (RG) limit cycle within Short-Range Effective Field Theory (SREFT).  While the analytic form of the three-body renormalization relation has been established for a sharp cutoff regulator, its universality for other regulators remains underexplored.  In this work, we derive the universal functional form of the three-body renormalization relation for general separable regulators through a detailed analysis of the Skorniakov-Ter-Martirosian and Faddeev equations.  We find that the relation follows from a real M\"{o}bius transformation characterized by three parameters.  This universality is verified numerically for various regulators.  Although the functional form remains the same, the parameters characterizing the limit cycle exhibit regulator dependence.  These findings broaden the class of RG limit cycles in SREFT and offer a more complete understanding of three-body renormalization.
\end{abstract}

\maketitle

\section{Introduction}
The concept of a limit cycle was
introduced by Wilson as a special solution to the renormalization group (RG) equations in the early development of RG theory~\cite{Wilson:1973jj}.  A limit cycle in the RG flow implies that physical observables recur periodically under scale transformations, reflecting discrete scale invariance (DSI) rather than continuous scale invariance.  A prominent physical realization of an RG limit cycle is found in the three-body problem with short-range interactions.  In his seminal work~\cite{Efimov:1970zz}, Efimov showed that three identical bosons with resonant two-body interactions exhibit an infinite tower of bound states whose energies form a geometric sequence, a hallmark of DSI. This phenomenon, known as the Efimov effect, has attracted substantial experimental interest \cite{Ferlaino:2009zz, Zaccanti:2009ugr, Kunitski:2015qth} and was later identified as a concrete manifestation of an RG limit cycle within the framework of Short-Range Effective Field Theory (SREFT)~\cite{Bedaque:1998kg, Bedaque:1998km}.  Further developments in this direction have deepened our understanding of the Efimov effect and extended the implications of DSI to more complex few- and many-body systems~\cite{Braaten:2004rn, vanKolck:2017jon}, which has profound implications for systems in atomic, nuclear, condensed matter, and particle 
physics~\cite{Nielsen:2001hbm,Braaten:2006vd,Hammer:2010kp,Nishida:2012hf,Pal:2012ioz,Naidon:2016dpf,Kievsky:2021ghz}.

SREFT captures the most general short-range dynamics allowed by assumed spacetime symmetries.  Its leading order (LO) Lagrangian density reads
\begin{equation}
    \mathcal{L} = \psi^{\dagger}\left( i \partial_0 + \frac{\vec{\nabla}^2}{2m} \right) \psi - \frac{C_0}{2} \left( \psi^{\dagger} \psi\right)^2
    -\frac{D_0}{6} \left( \psi^{\dagger} \psi\right)^3
    , \;\;
    \label{L_SREFT}
\end{equation}
where $\psi$ is a spinless particle field, $m$ is the particle mass, and $C_0(\Lambda)$ and $D_0(\Lambda)$ are the two- and three-body low-energy constants (LECs), respectively.  Their running with the regulator scale $\Lambda$, known as the renormalization relation, is determined by two physical observables: the two-body scattering length and the three-body parameter.  In the unitarity limit, where the two-body scattering length diverges, there is no dimensionful parameter in the two-body system at LO and $C_0(\Lambda)$ has a universal dependence on $\Lambda$, $mC_0(\Lambda)\propto \Lambda^{-1}$, corresponding to a nontrivial RG fixed point~\cite{Weinberg:1991um}.  Denoting the dimensionful three-body parameter by $\Lambda_\ast$, dimensional analysis requires $mD_0(\Lambda_\ast, \Lambda)\propto \Lambda^{-4}$, with the dimensionless ratio
\begin{equation}
    H_0(\Lambda/\Lambda_\ast) \equiv \frac{\Lambda^2 D_0(\Lambda_\ast, \Lambda)}{6 m C_0^2(\Lambda)}
    \equiv - \Lambda^2 h(\Lambda_\ast,\Lambda) 
    \label{H0_SREFT}
\end{equation}
a function of $\Lambda/\Lambda_\ast$.

The expression for $H_0(\Lambda/\Lambda_\ast)$ can depend on the specific choice of regulator.  With certain local regulators, it can even be multi-valued, corresponding to multiple branches of the limit cycle~\cite{Beane:2000wh, Bawin:2003dm, Kirscher:2015yda}.  In this work, we restrict our analysis to separable (and thus nonlocal) regulators, which yield a unique and well-defined branch, and are widely employed in few- and many-body calculations. It has been shown that for a sharp momentum cutoff $H_0$ has the analytical form~\cite{Bedaque:1998kg, Bedaque:1998km, Braaten:2011sz, Chen:2025rti}
\begin{equation}
    H_0(\Lambda/\Lambda_\ast) = h_0 \frac{ \sin(s_0 \ln (\Lambda/\Lambda_\ast) - \delta_0 ) }{ \sin(s_0 \ln (\Lambda/\Lambda_\ast) + \delta_0 ) }\;,
    \label{H0_0_expr}
\end{equation}
where $s_0 \simeq$ 1.00624 is a universal constant characteristic of the limit cycle, and $h_0$ and $\delta_0$ are pure numbers.  DSI implies a geometric tower of bound states with binding momenta $\kappa_\ast \exp({-}l\pi/s_0)$, where $l$ is an integer and $\kappa_{\ast} \equiv \sqrt{m B_3}$ corresponds to a reference state ($l=0$) with binding energy $B_3$.  Again by dimensional analysis,  
$\kappa_{\ast}$ is related to $\Lambda_\ast$ by a pure number 
\begin{equation}
b_0\equiv \Lambda_\ast/\kappa_{\ast}.
\label{eq:b0}
\end{equation}

For the special case of a sharp cutoff regulator in the Skorniakov–Ter-Martirosian (STM) equation~\cite{Skorniakov:1957kgi}, the phase $\delta_0=\arctan(s_0^{-1})$~\cite{Bedaque:1998kg, Bedaque:1998km},
and expressions for both $h_0$ and $b_0$ 
have been obtained recently in terms of certain integrals~\cite{Chen:2025rti}. 
The renormalization relation for other regulators  and different dynamical equations has been less explored. 
The form~\eqref{H0_0_expr}, with the same fixed phase $\delta_0=\arctan(s_0^{{-}1})$, has been used as a fitting formula for a separable Gaussian regulator~\cite{Platter:2004zs, Platter:2005sj}.

In this work, we demonstrate that Eq.~\eqref{H0_0_expr} represents a universal functional form for general separable regulators through a detailed analysis of the STM and Faddeev~\cite{Faddeev:1960su} equations.  This universality is explicitly confirmed by numerical simulations using various regulators.  Our results further reveal that $b_0$, $h_0$, and $\delta_0$ depend on the specific choice of regulator, indicating that Eq.~\eqref{H0_0_expr} represents a broader class of RG limit cycles beyond the commonly adopted form with fixed phase.

The paper is organized as follows. In Sec.~\ref{sec:STM}, we derive Eq.~\eqref{H0_0_expr} based on the STM equation. A complementary derivation starting from the Faddeev equation is provided in Sec.~\ref{sec:Faddeev}. Section~\ref{sec:numerical} presents our numerical results for various separable regulators. Approximate expressions for the parameters characterizing the limit cycle in the case of (super-)Gaussian regulators in the STM equation are given in Sec.~\ref{sec:approx}.  The structure of the newly derived form of the limit cycle is studied in Sec.~\ref{sec:structure}. We conclude in Sec.~\ref{sec:conclusion}.

\section{STM Equation with General Regulator} 
\label{sec:STM}
We begin with the theoretical analysis of the three-body problem in SREFT for general separable regulators. 
This analysis can be simplified by introducing an auxiliary dimer field $d$, which allows the Lagrangian density to be reformulated as~\cite{Bedaque:1998kg,Braaten:2004rn}
\begin{multline}
    \mathcal{L} = \psi^{\dagger}\left( i \partial_0 + \frac{\mathbf{\nabla}^2}{2m} \right) \psi 
    + \frac{d^\dagger d}{2m C_0} 
    \\
    - \frac{1}{2\sqrt{m}} \left( d^\dagger \psi \psi + d \psi^\dagger\psi^\dagger \right)
    + h d^\dagger d \psi^\dagger \psi + \ldots \;,
    \label{L_dimer}
\end{multline}
with $h$ defined in Eq.~\eqref{H0_SREFT}.

\begin{figure}[t!]
    \centering
    \includegraphics[width=0.85\linewidth]{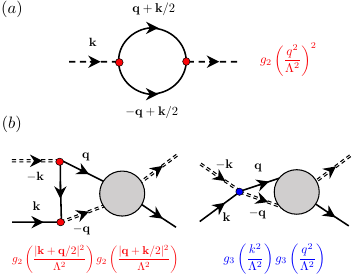}
    \caption{(a) Self-energy diagram of the dimer field $d$. Solid and dashed lines represent the propagators of the particle $\psi$ and the dimer $d$, respectively.  (b) Scattering process between a dimer $d$ and a particle $\psi$.  The double-dashed lines indicate the renormalized propagator of the dimer field $d$, which can depend on the two-body regulator $g_2$.  Only the momenta of the fields are labeled, since they correspond to the arguments of the two-body and three-body regulators, $g_2$ and $g_3$.}
    \label{fig:diagrams}
\end{figure}

The three-body problem reduces to the scattering between a boson $\psi$ and a dimer $d$.  This requires 
the renormalized propagator for a dimer of energy $E$ and momentum $\mathbf{k}$, $D_r(E, \mathbf{k})= [(2mC_0)^{-1}-\Sigma(E,\mathbf{k})]^{{-}1}$.  The self-energy 
\begin{equation}
\Sigma(E,\mathbf{k})=\frac{1}{2m}\int\!\!\frac{d^3
q}{(2\pi)^3}\frac{g_2^2(q^2/\Lambda^2)}{E_+ - \epsilon_{\mathbf{q} - \mathbf{k}/2} - \epsilon_{\mathbf{q}+\mathbf{k}/2}}\;,
\label{eq:Sigma}
\end{equation}
where $E_+ = E + i0^+$ and $\epsilon_\mathbf{k}=k^2/2m$, receives contributions only from the diagram shown in Fig.~\ref{fig:diagrams}(a).  Here, $g_2(x)$ is the two-body regulator that ensures the convergence of the above integral: a function satisfying the asymptotic behavior $g_2(x)=1$ for $x\rightarrow0$ and $g_2(x)=0$ for $x\rightarrow \infty$.  By tuning $C_0=(2m \Sigma(0,\mathbf{0}))^{-1}$, the system reaches the two-body unitarity limit, and the propagator becomes
\begin{equation}
D_r(E,\mathbf{k})^{{-}1} = {-}\frac{1}{8\pi}\sqrt{{-}mE_{r}}\,\chi({-}mE_r/\Lambda^2),
\label{D_r}
\end{equation}
where $E_r = E_+ - k^2/4m$ is the energy in the center-of-mass frame.  The regulator-dependent function $\chi(x)$ serves as a high-energy correction to the dimer propagator: when the relevant momentum scale is much smaller than the cutoff $\Lambda$ ($x \ll 1$), $\chi(x)$ approaches 1, and the corrections are suppressed by inverse powers of $\Lambda$, which can be dropped at LO.  One then recovers the expected low-energy $T$ matrix for two-body scattering.

The relevant diagrams for the three-body problem are shown in Fig.~\ref{fig:diagrams}(b).  We focus on the bound-state sector with total energy $E < 0$ and zero total momentum.  In this case, the $s$-wave bound-state wave function $\phi(k)$ satisfies the STM equation with general separable regulators,
\begin{equation}\label{eq:STM_general_reg}
\begin{aligned}
    k\phi(k)=&-\!\int\!\!\frac{dq
    }{2\pi^2} \left( G_r(k,q;E)
    -\frac{kq}{\Lambda^2} \, H_0(\Lambda) \right.\\
    &\left. \null \times 
    g_3\!\left(\frac{k^2}{\Lambda^2}\right)
    g_3\!\left(\frac{q^2}{\Lambda^2}\right)\right)
    D_r(E-\epsilon_{\mathbf q},-\mathbf{q})\, q\phi(q) \;,
    \end{aligned}
\end{equation}
where
\begin{equation}\label{eq:Grdef}
G_r(k,q;E)\equiv kq\int \! \frac{d\Omega_\mathbf{q}}{4\pi}\frac{ g_2\left(\frac{|\mathbf{k}+\mathbf{q}/2|^2}{\Lambda^2}\right) g_2\left(\frac{|\mathbf{q}+\mathbf{k}/2|^2}{\Lambda^2}\right)}{m(\epsilon_{\mathbf k}+\epsilon_{\mathbf q}+\epsilon_{\mathbf{k}+\mathbf{q}}-E)}
\end{equation}
involves an average over the momentum-space angles.  Here, $g_3(x)$ is a separable three-body regulator that shares the same asymptotic behavior as $g_2(x)$, but can take a different functional form. 

We focus on shallow bound states with ${-}mE\ll \Lambda^2$.  In the intermediate regime with $\sqrt{{-}mE} \ll k \ll \Lambda$, the solution of Eq.~\eqref{eq:STM_general_reg} can be analyzed by taking $E \rightarrow 0$ and $\Lambda \rightarrow \infty$.  This corresponds to setting $\chi(x)=g_2(x)=g_3(x)=1$ and $h=0$, where the equation becomes scale invariant and the solution $\phi(k) \sim k^{{-}1} \cos\left(s_0 \ln (k/\tilde{\Lambda}_\ast)\right)$ exhibits DSI. The parameter $\tilde{\Lambda}_\ast\equiv
\exp(\varphi_0/s_0) \Lambda_\ast$ depends on the regularization scheme.  An integral expression for $\varphi_0$ with a sharp regulator can be found in 
Ref.~\cite{Chen:2025rti}.
 
We are now ready to derive the universal functional form~\eqref{H0_0_expr} for generic regulators $g_2(x)$ and $g_3(x)$, where $\Lambda_\ast$, $\delta_0$, and $h_0$ are treated as fit parameters.  Focusing on the limit $E \rightarrow 0$ and introducing $k \equiv \Lambda \exp({-}t)$, $q \equiv \Lambda \exp(-s)$, $\xi(t)\equiv k\phi(k)$, $\tilde{g}_3(t)\equiv \exp(-t)\,g_3(\exp(-2t))$,
$\tilde{\chi}_r(t)\equiv\chi(3\exp(-2t)/4)$, and $\lambda \equiv \sqrt{3}\pi/8$,
the STM equation can be recast as
\begin{equation}
\label{eq:STM_general_t}
    \int_{{-}\infty}^{\infty} \! ds \left[G_r(t,s)-H_0 \, \tilde{g}_3(t)\, \tilde{g}_3(s)\right]\tilde{\chi}^{{-}1}_r(s)\xi(s)=\lambda\xi(t) \;.
\end{equation}
The function $G_r(t, s)=G_r(k,q;0)$ approaches
\begin{equation}
    G(t-s) \equiv \frac{1}{2} \ln \left( \frac{\cosh(t-s)+1/2}{\cosh(t-s)-1/2} \right),
\end{equation}
if both $s$ and $t$ are large and positive.  If either $t \ll 0$ or $s \ll 0$, $G_r(t,s)$ goes to 0. 

To eliminate $\tilde{\chi}_r(t)$ in Eq.~\eqref{eq:STM_general_t}, we make the redefinitions
\begin{equation}
\begin{aligned}
    &B(t,s)\equiv G_r(t,s)/\sqrt{\tilde{\chi}_r(t)\tilde{\chi}_r(s)} \;,\\
    &\psi(t)\equiv \xi(t)/\sqrt{\tilde{\chi}_r(t)}, \quad v(t)\equiv \tilde{g}_3(t)/\sqrt{\tilde{\chi}_r(t)} \;,
    \end{aligned}
\end{equation}
which allow us to rewrite the equation as
\begin{equation}
    \int_{{-}\infty}^{\infty}\! ds\left[B(t,s)-H_0v(t)v(s)\right]\psi(s)=\lambda\psi(t) \;.
    \label{eq:STM_proof}
\end{equation}
Because the asymptotic properties of $B(t,s)$ are the same as those of $G_{r}(t,s)$, the asymptotic behavior of $\psi(t)$ is the same as that for a sharp cutoff regulator \cite{Chen:2025rti},
\begin{equation}
\label{eq:asymptotic}
 \psi(t) \sim \cos(s_0 t+\tilde{\varphi}) \;,
\end{equation}
with $\tilde{\varphi} = s_0\ln[\tilde{\Lambda}_\ast/\Lambda]$. Determining the running of $H_0$ is equivalent to analyzing how the phase of the solution $\psi(k)$ depends on $H_0$ in Eq.~\eqref{eq:STM_proof}.

Since Eq.~\eqref{eq:STM_proof} is homogeneous, the overall normalization of $\psi$ can be chosen freely. 
With
\begin{equation}
    \int_{{-}\infty}^{\infty} \! ds \, v(s)\psi(s)=1 \;,
    \label{psi_norm}
\end{equation}
one obtains the inhomogeneous equation
\begin{equation}
    \int_{-\infty}^{\infty}\! ds\left[B(t,s)- \lambda \delta(s-t)\right]\psi(s)=H_0v(t) \;.
    \label{eq:STM_proof2}
\end{equation}
We emphasize that the removal of $\psi$ from the second term in Eq. \eqref{eq:STM_proof} is of vital importance, 
and only possible when the three-body regulator is separable.  We denote by $\psi_0(s)$ and $\psi_1(s)$ the solutions of Eq.~\eqref{eq:STM_proof2} for $H_0 = 0$ and $H_0 = 1$, respectively, that satisfy Eq.~\eqref{psi_norm}.  The general solution can be written as the sum of the solution of the homogeneous equation and a particular solution of the inhomogeneous equation,
\begin{equation}
    \psi(t) = (1-H_0) \psi_0(t) + H_0 \psi_1(t) \;,
    \label{psi_sol}
\end{equation}
which can be verified by directly substituting Eq.~\eqref{psi_sol} into Eq.~\eqref{eq:STM_proof2} and using the definitions of $\psi_0(s)$ and $\psi_1(s)$.  The coefficient in front of $\psi_0(t)$ is fixed by the normalization condition in Eq.~\eqref{psi_norm}. This solution depends linearly on $H_0$.

In the low-energy regime $t\gg 1$, both $\psi_0(t)$ and $\psi_1(t)$ are expected to exhibit periodic behavior with frequency $s_0$: $\psi_0(t)\sim \text{Re}\!\left[A_0 e^{is_0 t}\right]$ and $\psi_1(t)\sim \text{Re}\!\left[(A_0+A_1) e^{is_0 t}\right]$, where $A_0$ and $A_1$ are complex numbers. Matching with Eq.~\eqref{eq:asymptotic} gives
\begin{equation}
    \tan\tilde\varphi=\tan\left(\arg(A_0+H_0A_1)\right)
    = \frac{\text{Im}(A_{0})+\text{Im}(A_{1})H_0 }{\text{Re}(A_{0})+\text{Re}(A_{1})H_0} \;.
    \label{tanvarphi}
\end{equation}
The mapping from $H_0$ to $\tan\tilde\varphi$ is therefore a real M\"{o}bius transformation, which forms a group under composition and involves three independent parameters.
In particular, $A_0$ and $A_1$ are determined only up to a common overall rescaling. Conversely, $H_0$ can be expressed as a real M\"{o}bius transformation of $\tan \tilde{\varphi}$. An arbitrary parametrization of this relation may be employed. Choosing $\arg(A_0 A_1)=2\varphi_0$ and setting 
$\text{Re}[\exp({-}i\varphi_0) A_1] =1$, we can write
\begin{equation}
\begin{aligned}
    A_0 &= {-}h_0 \left(1- i \tan \delta_0\right) e^{i\varphi_0} \;, \\
    A_1 &= \left(1 + i\tan \delta_0\right) e^{i\varphi_0} \;,
\end{aligned}
\end{equation}
where $\delta_0$ and $h_0$ are real and can be determined numerically.
From Eq.~\eqref{tanvarphi} we obtain
\begin{equation}
H_0= h_0 \frac{\tan(\varphi_0-\tilde\varphi)-\tan \delta_0}{\tan(\varphi_0-\tilde\varphi)+\tan \delta_0} \,,
\label{H0_0_res}
\end{equation}
which is equivalent to Eq.~\eqref{H0_0_expr}.

\section{Faddeev Formalism}
\label{sec:Faddeev}
An alternative description of the three-body problem is provided by the Faddeev equation, built upon the Hamiltonian formulation.  We show that the general form~\eqref{H0_0_expr} also holds for this formulation with separable two- and three-body potentials,
described in detail in Ref.~\cite{Konig:2019xxk}.

In terms of Jacobi momenta $\mathbf{u}_1=\left(\mathbf{p}_1 - \mathbf{p}_2 \right)/2$ and $\mathbf{u}_2= 
2\left[\mathbf{p}_3 -
\left(\mathbf{p}_1 + \mathbf{p}_2 \right)/2 \right]/3$, where $\mathbf{p}_i$
is the momentum of the $i$th particle, the two-body potential for 
the first pair reads
\begin{equation}
 \braket{
    \mathbf{u}_1'| V_2 | \mathbf{u_1}}
    = C_0 \braket{\mathbf{u}_1'|g_2}
    \braket{g_2|\mathbf{u}_1}
    \label{V2_0}
\end{equation}
with $\braket{\mathbf{u}_1 | g_2} = g_2(u_1^2/\Lambda^{2})$.  Similarly, the three-body potential is
\begin{equation}
    \braket{ \mathbf{u}_1' \mathbf{u}_2'
    | V_3 | \mathbf{u}_1 \mathbf{u}_2
    } = D_0  \braket{ \mathbf{u}_1' \mathbf{u}_2'
    | \zeta } \braket{ \zeta | \mathbf{u}_1 \mathbf{u}_2
    }
    \label{V3}
\end{equation}
with
    $\braket{ \mathbf{u}_1 \mathbf{u}_2
    | \zeta } \equiv \zeta(u_1,u_2)= \zeta\! \left((u_1^2 + \tfrac34 u_2^2)/\Lambda^2\right)$. 
With $P=P_{12}P_{23} + P_{13}P_{23}$ generating cyclic and anti-cyclic permutations, the Faddeev equation for one of three equivalent Faddeev amplitudes $\ket{\uppsi}$ can be written as
\begin{equation}
 \ket{\uppsi} = G_0 t P\ket{\uppsi} + 3 G_0 t G_0 t_3 \ket{\uppsi} \,.
\label{eq:Faddeev-symm}
\end{equation}
Here, the free three-body Green’s function $G_0$, and the 
two- and three-body $T$ matrices $t$ and $t_3$  implicitly depend on the energy $E$. Solving the Lippmann-Schwinger equation for
the potential~\eqref{V2_0} with $C_0$ tuned to unitarity gives 
    $t(z) = \ket{g_2} \tau(z) \bra{g_2}$,
where $\tau(z)={D_r(z,\mathbf{0})}/{2m}$ with $D_r(z,\mathbf{0})$ in Eq.~\eqref{D_r}. The Lippmann-Schwinger-like equation $t_3 = V_3 + V_3 G_0 t_3$ can be solved
algebraically for the separable three-body potential that we use, giving
$t_3(E) = \ket{\zeta} \tau_3(E) \bra{\zeta}$, where
$\tau_3(E) = {-}H_0'(E,\Lambda)/I_2^{\zeta}(E)$ with
$I_2^{\zeta}(E) = \braket{\zeta| G_0(E) | \zeta}$ and
\begin{equation}
    H_0'(E,\Lambda) = \frac{H_0(\Lambda)}
    {H_0(\Lambda)-\Lambda^2/6m C_0^2 I_2^{\zeta}(E)} \,.
\end{equation}

Since the Faddeev equations are closed when the interaction is active only in certain partial waves, we only need the total $s$-wave contribution for the three-boson system ($l_1 = l_2 = 0$ for the orbital angular momenta $l_{1,2}$ associated with ${\mathbf u}_{1,2}$). Starting from Eq.~\eqref{eq:Faddeev-symm}, we obtain
\begin{equation}
\begin{aligned}
    \braket{\mathbf{u}_1 \mathbf{u}_2 | \uppsi} =& G_0(E; u_1, u_2) \braket{\mathbf{u}_1 \mathbf{u}_2 | t P | \uppsi} \\
    &+ 3 G_0(E; u_1, u_2) \braket{\mathbf{u}_1 \mathbf{u}_2 |t G_0 t_3 | \uppsi} \;.
\end{aligned}
\end{equation}
Inserting the separable forms of the $T$ matrices $t(z)=\ket{g_2} \tau(z) \bra{g_2}$ and $t_3(E)=\ket{\zeta} \tau_3(E) \bra{\zeta}$, we get
\begin{widetext}
\begin{multline}
    \uppsi(u_1, u_2) = 2 G_0(E; u_1, u_2) \, g_2\!\left(\frac{u_1^2}{\Lambda^2}\right) \tau\!\left(E-\frac{3u_2^2}{4m} \right) \int \!\!\frac{d^3 u_2'}{(2\pi)^3} \, g_2\!\left(\frac{|\bm{\pi}(\mathbf{u}_2,\mathbf{u}_2')|^2}{\Lambda^2} \right) \uppsi\!\left( \pi(\mathbf{u}_2',\mathbf{u}_2), u_2' \right) \\
    \null + 3\,  G_0(E; u_1, u_2)  g_2\!\left(\frac{u_1^2}{\Lambda^2}\right) \tau\!\left(E-\frac{3u_2^2}{4m} \right) I_0^{\zeta}(E,u_2) \,\tau_3(E) \braket{\zeta| \uppsi}\;,
\label{uppsi_u1_u2}
\end{multline}
in which $\uppsi(u_1, u_2) = \braket{\mathbf{u}_1 \mathbf{u}_2 | \uppsi}$ since there is no angular dependence for the 
$s$ wave,
\begin{equation}
    \bm{\pi}(\mathbf{u}_2,\mathbf{u}_2') = \mathbf{u}_2/2+\mathbf{u}_2' \;,
\end{equation}
\begin{equation}
    I_0^{\zeta}(E,u_2) = \int\!\!\frac{d^3 u_1}{(2\pi)^3} \, g_2\!\left(\frac{u_1^2}{\Lambda^2} \right) G_0(E; u_1,u_2) \, \zeta(u_1,u_2) \;,
    \label{eq:I0_zeta}
\end{equation}
and $\braket{\zeta| \uppsi}$ is a number given by another integral.

Following Refs.~\cite{Mitra:1962three, Platter:2005}, we define a reduced Faddeev component $F(u_2)$ via
\begin{equation}
 \uppsi(u_1, u_2) 
 = g_2(u_1^2/\Lambda^2) \, G_0(E;u_1,u_2) \, \tau\!\left(E-\tfrac34u_2^2\right) F(u_2) \;,
\label{eq:psi-to-F}
\end{equation}
which satisfies
\begin{multline}
    F(u_2) = 2 \int \!\! \frac{d^3 u_2'}{(2\pi)^3} \, g_2\!\left(\frac{|\bm{\pi}(\mathbf{u}_2,\mathbf{u}_2')|^2}{\Lambda^2} \right)
    g_2\!\left(\frac{|\bm{\pi}(\mathbf{u}_2',\mathbf{u}_2)|^2}{\Lambda^2} \right) G_0\!\left(E; \mathbf{\pi}(\mathbf{u}_2',\mathbf{u}_2\right), \mathbf{u}_2') \, \tau\!\left(E-\frac{3 u_2'^2}{4m}\right) F(u_2') \\
    \null + 3 I_0^{\zeta}(E,u_2) \tau_3(E) \, \int \!\! \frac{d^3 u_2'}{(2\pi)^3} \, I_0^{\zeta}(E,u_2') \, \tau\!\left(E-\frac{3 u_2'^2}{4m} \right) F(u_2') \;.
    \label{eq:F_u2_Faddeev}
\end{multline}
With $\tau(z)={D_r(z,\mathbf{0})}/{2m}$ and $\tau_3(E) = {-}H_0'(E,\Lambda)/I_2^{\zeta}(E)$, this can be brought into the following form 
\begin{equation}\label{eq:Faddeev_general_reg}
\begin{aligned}
u_2F(u_2)={-}\!\int\!\!\frac{d u_2'}{2\pi^2}& \left(G_r(u_2, u_2', E)-\frac{u_2 u_2'}{\Lambda^2} \, H_0'(E,\Lambda) \, g_3'\!\left(E,u_2\right) g_3'\!\left(E,u_2'\right)\right)
D_r(E-\epsilon_{\mathbf u_2'},-\mathbf{u}_2')\, u_2' F(u_2') \;,
\end{aligned}
\end{equation} 
\end{widetext}
where
\begin{equation}
    g_3'(E, u_2) = {-}\sqrt{3}\Lambda I_0^{\zeta}(E, u_2) /{\sqrt{-2 m I_2^{\zeta}(E)}}.
    \label{g3_p}
\end{equation}
In the special case where the three-body regulator factorizes as a function of the momenta, $\ket{\zeta} = \ket{g_2} \ket{g_3}$ in abstract notation, we have
\begin{equation}
    g_3'(E,u_2) =  {-}\frac{ \sqrt{6 m}\Lambda \Sigma(E-3u_2^2/4m, \mathbf{0})}{\sqrt{{-} I_2^{\zeta}(E)}} \, g_3\!\left( \frac{u_2^2}{\Lambda^2}\right) \;.
    \label{g2p_sep}
\end{equation}

Equation~\eqref{eq:Faddeev_general_reg} shows that $F(u_2)$ corresponds to $\phi(k)$ in the absence of a three-body force.  It has the same structure of the STM equation \eqref{eq:STM_general_reg},  
only $g_3'$ and $H_0'$ are energy dependent.  The derivation based on the STM equation, which takes $E\to0$, can be adapted to Eq.~\eqref{eq:Faddeev_general_reg} with only minor modifications.  This leads to a similar asymptotic behavior and a linear dependence of $F(u_2)$ on $H_0'(0,\Lambda)$, which is related to $H_0$ through another real M\"{o}bius transformation. Eq.~\eqref{H0_0_expr} therefore remains valid.

Using the fact that $(1-G_0 V_3)^{{-}1} = 1+G_0 t_3$ and the total symmetry of $G_0$ and $V_3$ under particle exchange, the Faddeev equation~\eqref{eq:Faddeev-symm} can be transformed into the equivalent form
\begin{equation}
 \ket{\tilde{\uppsi}} = G_0 t P\ket{\tilde{\uppsi}} + (G_0 + G_0 t G_0) V_3 \ket{\tilde{\uppsi}} \,,
\label{eq:Faddeev-V3}
\end{equation}
where $\ket{\tilde{\uppsi}}$ is one of three equivalent two-body Faddeev components,
related to the Faddeev component $\ket{\uppsi}$ via
\begin{equation}
    \ket{\uppsi} = (1- G_0 V_3) \ket{\tilde{\uppsi}} \;.
    \label{psi_psi_tilde}
\end{equation}
For a derivation of this form directly from the Schr{\"o}dinger equation, see Ref.~\cite{Stadler:1991zz}. Instead of Eq.~\eqref{eq:Faddeev-symm}, one could also start from Eq.~\eqref{eq:Faddeev-V3} and use the relation in Eq.~\eqref{psi_psi_tilde} to obtain Eq. \eqref{eq:Faddeev_general_reg}.
For the numerical solution of the Faddeev equation, we use Eq.~\eqref{eq:Faddeev-V3}.

\section{Numerical Demonstration}
\label{sec:numerical}
To validate our analytical prediction, we apply Eq.~\eqref{H0_0_expr} to 
results from the numerical solution 
of both the STM equation~\eqref{eq:STM_general_reg} and the Faddeev equation~\eqref{eq:Faddeev-V3}.  Following standard practice in the literature, we set $\chi=1$ in Eq.~\eqref{D_r} for the STM calculation,
while retaining it in the Faddeev formalism: 
$\chi$ affects the numerical values of the parameters in Eq.~\eqref{H0_0_expr}, but not 
our main conclusion. 

\begin{table}[!tb]
\setlength{\extrarowheight}{1.2pt}
\caption{Values of the dimensionless parameters appearing in Eqs.~\eqref{H0_0_expr} and \eqref{eq:b0} for various regulators: sharp cutoff from the literature; and Gaussian ($n=1$), quartic super-Gaussian ($n=2$), and sextic super-Gaussian ($n=3$), as determined in this work by solving the STM and Faddeev equations.}
\begin{center}
\begin{tabular}{ c c c c c}
\hline
\hline
 & regulator& $\delta_0$ & $h_0$ & $b_0$ \\
 \hline
 \multirow{4}{*}{STM} &sharp & 0.7823 \cite{Bedaque:1998kg,Bedaque:1998km} & 0.879 \cite{Braaten:2011sz} & 2.61 \cite{Braaten:2004rn,Ji:2015hha} \\
 &$n=1$ & 1.0463 & 1.8024 & 4.4436 \\
 &$n=2$ & 0.8869 & 1.4744 & 3.4930 \\
 &$n=3$ & 0.8361 & 1.2804 & 3.2042 \\
 \hline
 \multirow{3}{*}{Faddeev}& $n=1$ & 0.7094 & 0.7976 & 2.3965 \\
 &$n=2$ & 0.4455 & 1.0189 & 2.6236 \\
 &$n=3$ & 0.3766 & 1.0037 & 2.5985 \\
 \hline
 \hline
\end{tabular}
\end{center}
\label{H0paramvalues}
\end{table}

\begin{figure*}[!tb]
		\centering
		\includegraphics[width=0.95\columnwidth]{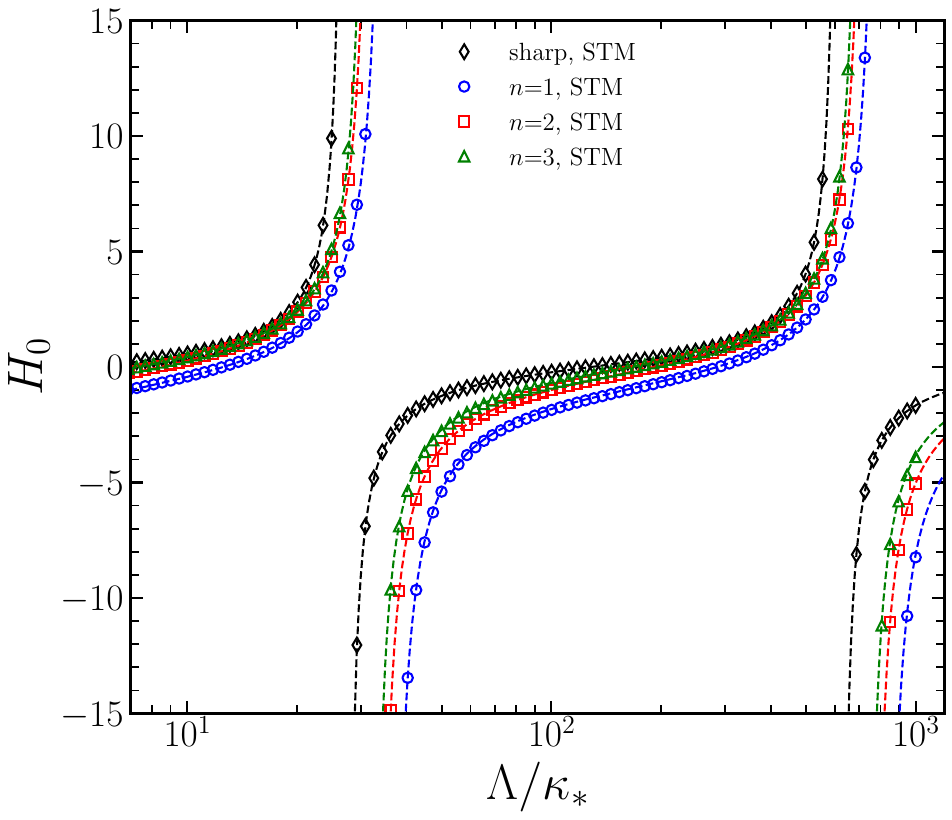}\hfil
		\includegraphics[width=0.95\columnwidth]{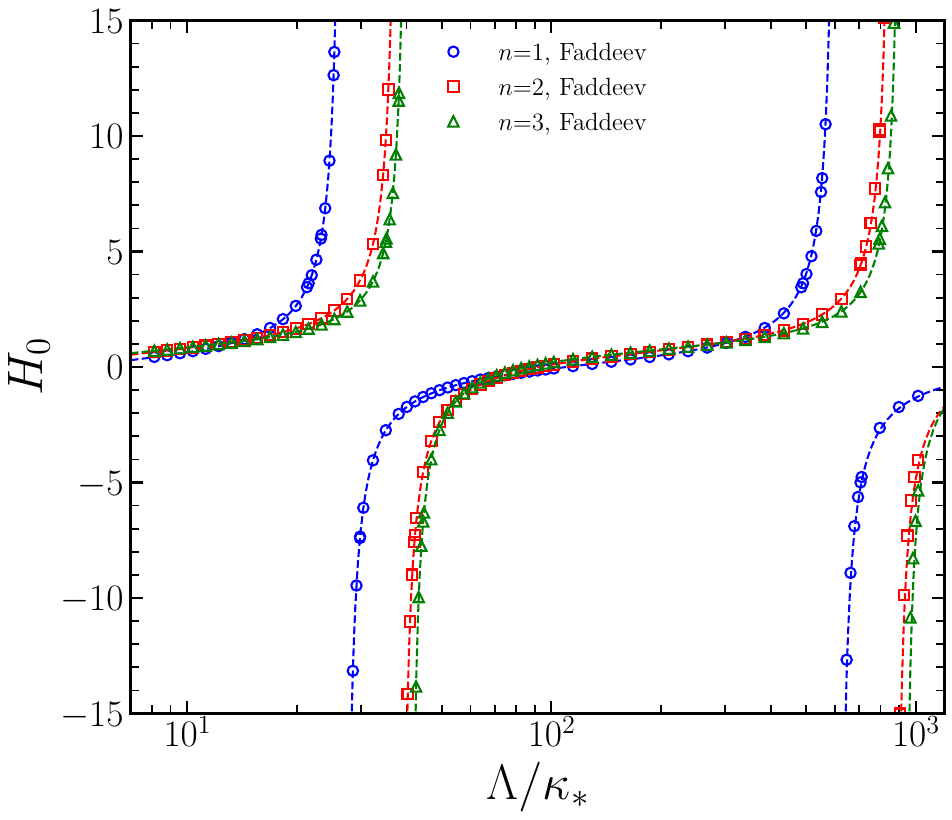}
		\vspace{-1em}
		\caption{
        The three-body low-energy constant $H_0$ at the unitarity limit as a function of the momentum cutoff $\Lambda$ (in units of $\kappa_{*}$) for the sharp cutoff (black diamonds), Gaussian ($n=1$, blue circles), quartic super-Gaussian ($n=2$, red squares), and sextic super-Gaussian ($n=3$, green triangles). Results obtained from solving the STM and Faddeev equations are shown in the left and right panels, respectively. The black dashed line is obtained with parameter values from the literature for the sharp cutoff (see Table \ref{H0paramvalues}). The lines for (super-)Gaussians are obtained by fitting the data points of the same color with Eq. \eqref{H0_0_expr}. 
        }
	\label{fig:H0}
\end{figure*}

In addition to the sharp momentum cutoff, we also consider smoother regulator functions
\begin{equation}
    g_2(x^2)=g_3(x^2) = \zeta(x^2)
    = \exp(-x^{2n}) \;,
\label{eq:SuperGaussians}
\end{equation}
where $n=$ 1, 2, 3, $\ldots$ correspond, respectively, to the standard, quartic, sextic, \textit{etc.}\ (super-)Gaussians. The special scenario for Eq. \eqref{g2p_sep} is realized for the simple Gaussian regulator, but not for the super-Gaussian ($n \geq 2$) regulators. For the two-body $T$ matrix to attain the unitarity-limit form, the two-body LEC must be given by
\begin{equation}
    C_0(\Lambda) = \frac{1}{2m \Sigma(0,\mathbf{0})} = {-}\frac{4\pi}{m \theta_1 \Lambda} \;,
\label{C0_0}
\end{equation}
where $\theta_1$ is a regulator-dependent number and 
\begin{equation}
    \theta_1=
    \left(2^{1/2n}n\pi\right)^{-1}\, \Gamma\!\left(\frac{1}{2n}\right)
    \label{theta_n2} \;
\end{equation}
for the regulators in Eq.~\eqref{eq:SuperGaussians}.

Results for $H_0(\Lambda)$ with different regulators are shown in Fig.~\ref{fig:H0} with numerical values for the parameters $\{\delta_0, h_0, b_0\}$ given in Table~\ref{H0paramvalues}.  For a sharp cutoff in the STM equation, our numerical results agree remarkably well with the analytic expression~\eqref{H0_0_expr} using parameter values from the literature (see Table~\ref{H0paramvalues}), which include a fixed phase $\delta_0 =\arctan(s_0^{-1})$. For the (super-)Gaussian regulators, fits with Eq.~\eqref{H0_0_expr}, where the phase is no longer fixed, show excellent agreement with numerical calculations across all cases presented, reinforcing the validity of our analytical analysis. 

\section{Approximate form from the STM equation}
\label{sec:approx}
In Ref. \cite{Chen:2025rti}, the STM equation with a sharp cutoff is related to a generalized Wiener–Hopf integral equation, which enables exact analytical treatment. However, this approach cannot be applied to general regulators. Nevertheless, approximate expressions for $\delta_0$ and $h_0$ can be obtained by extending the analysis of Ref.~\cite{Bedaque:1998kg}. 

The three-body force is constructed such that the wave function (or equivalently the scattering amplitude if including an inhomogeneous term) converges at large cutoffs 
\begin{align}
    \lim_{\Lambda\to\infty} \frac{\partial \phi(k) }{\partial \ln \Lambda } \to 0 \;.
\end{align}
Applying the limits and derivatives to the STM equation
\eqref{eq:STM_general_reg} and formally solving for $h(\Lambda)$ gives
\begin{widetext}
\begin{equation}
    \label{eq:h_general_reg}
    h(\Lambda)\simeq \frac{\displaystyle\int\!\! d (\ln\Lambda) \left( 
    \frac{\partial}{\partial \ln \Lambda } \int\!\!\frac{d^3q}{(2\pi)^3}
    g_2(|\mathbf{k}+\mathbf{q}/2|^2/\Lambda^2)
    g_2(|\mathbf{q}+\mathbf{k}/2|^2/\Lambda^2)
    G_0 D_r \phi(q)\right)_{k\ll \Lambda}}{\displaystyle\int\!\! d(\ln \!\Lambda)
    \left(\frac{\partial}{\partial \ln \Lambda } \int\!\!\frac{d^3q}{(2\pi)^3}
    g_3(k^2/\Lambda^2) g_3(q^2/\Lambda^2)
    D_r \phi(q)\right)_{k\ll \Lambda}}\,, 
\end{equation}
\end{widetext}
where $\int \! d(\ln\Lambda)$ is the indefinite integral over $\ln\Lambda$ without the constant, which is eliminated by the boundary condition $\phi(k) \to 0$ at large $k$.  For regulators whose derivatives with respect to $\Lambda$ vanish at $q\ll \Lambda$, the integral over $q$ is dominated by $q \sim \Lambda$.  To explicitly evaluate the integrals, we need to choose the form of the regulators and also take the asymptotic expansion of the integrand at large $q$. For the (super-)Gaussian regulators in Eq. \eqref{eq:SuperGaussians}, the following asymptotic expressions at $q\sim\Lambda \gg k$ will be used:
\begin{align}
    g_2\!\left(\frac{|\mathbf{k}+\frac{\mathbf{q}}{2}|^2}{\Lambda^2} \right)
    g_2\!\left(\frac{|\mathbf{q}+\frac{\mathbf{k}}{2}|^2}{\Lambda^2} \right) &\sim e^{-\left(1+2^{-2n}\right)\left(\frac{q}{\Lambda}\right)^{2n}}\;, \label{g2_approx}
    \\
    g_3\!\left(\frac{k^2}{\Lambda^2}\right) g_3\!\left(\frac{q^2}{\Lambda^2}\right)&\sim e^{-\left(\frac{q}{\Lambda}\right)^{2n}}\;, \label{g3_approx}
    \\
    G_0 &\sim -\frac{1}{q^2} \;,
    \\
    D_r &\sim \frac{1}{q}\;,
\end{align}
where the prefactor of $D_r$ is dropped as we are only interested in the ratio in Eq.~\eqref{eq:h_general_reg}. 

\begin{figure*}[!tb]
    \centering
    \begin{minipage}{0.65\columnwidth}
        \centering
        \includegraphics[width=\linewidth]{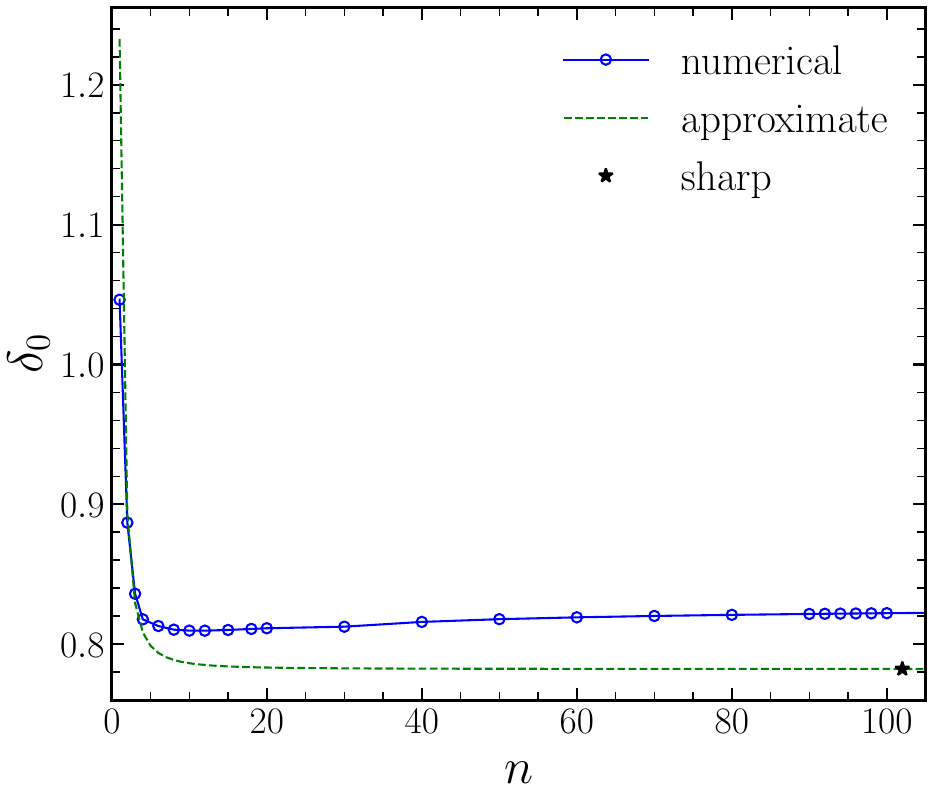}
        \par \qquad (a) 
    \end{minipage}
    \hspace{0.01\linewidth}
    \begin{minipage}{0.65\columnwidth}
        \centering
        \includegraphics[width=\linewidth]{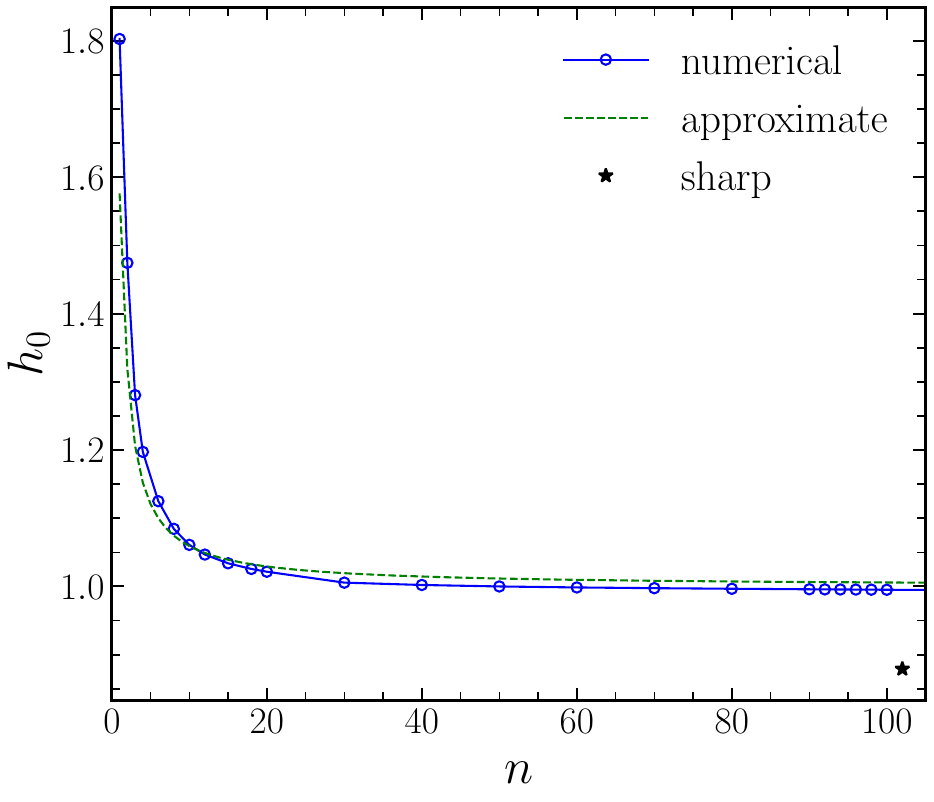}
        \par \qquad (b)
    \end{minipage}
    \hspace{0.01\linewidth}
    \begin{minipage}{0.65\columnwidth}
        \centering
        \includegraphics[width=\linewidth]{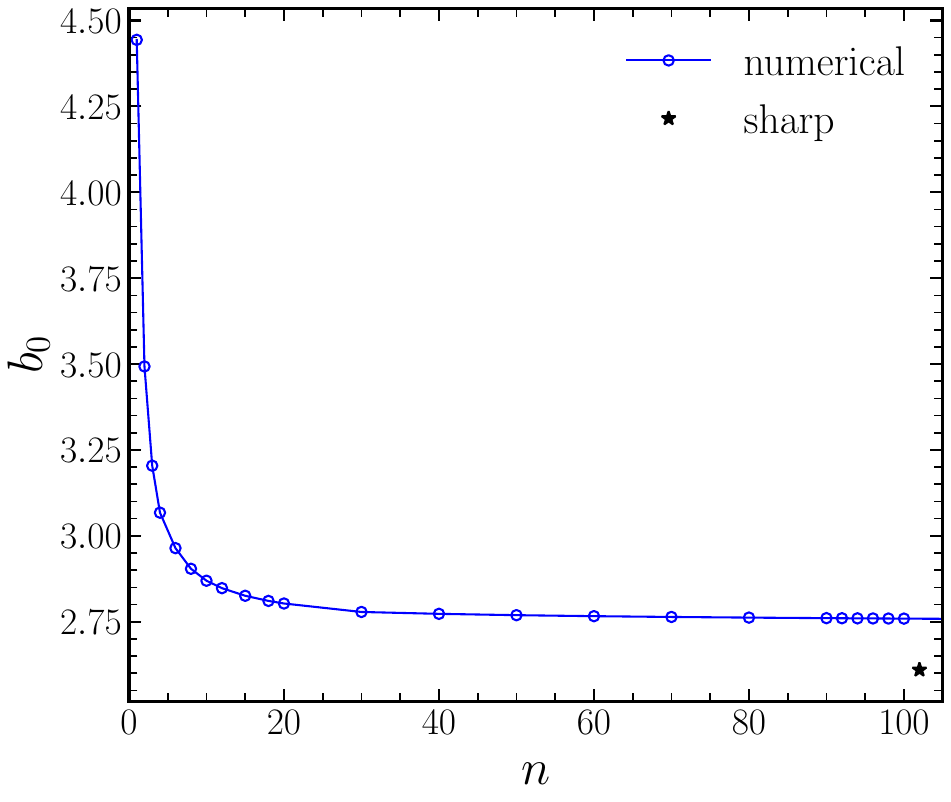}
        \par \qquad \;(c)
    \end{minipage}
    \caption{Dimensionless parameters $\delta_0$ (panel a), $h_0$ (panel b), and $b_0$ (panel c) appearing in Eqs.~\eqref{H0_0_expr} and \eqref{eq:b0} as functions of $n$, the index of a super-Gaussian regulator, Eq.~\eqref{eq:SuperGaussians}. Blue circles are obtained by numerically solving the STM equation~\eqref{eq:STM_general_reg} and fitting to Eq. \eqref{H0_0_expr}, green dashed lines represent the approximate values from Eqs. \eqref{eq:delta0approx} and \eqref{eq:h0approx}, and black stars denote the exact sharp-cutoff values from Ref.~\cite{Chen:2025rti}. }
    \label{fig:params_n}
\end{figure*}

We follow Ref.~\cite{Bedaque:1998kg} and take
\begin{equation}
    \phi(q) \sim \text{Re}\!\left[\left(\frac{q}{\tilde{\Lambda}_\ast}\right)^{is_0-1}\right] \;,
\end{equation}
as an approximation for $\phi(q)$ at $q \sim \Lambda$, although in principle this works only for $q \ll \Lambda$. Equation~\eqref{eq:h_general_reg} now becomes
\begin{equation}
\begin{aligned}
    \label{eq:h_general_reg_approx}
    h(\Lambda) &\simeq - \dfrac{\Re\displaystyle\int\!\!\dfrac{d^3q}{(2\pi)^3}
    e^{-a_n (q/\Lambda)^{2n}}
    q^{-3}(q/\tilde{\Lambda}_\ast)^{-1 + is_0}}{\Re\displaystyle\int\!\!\dfrac{d^3q}{(2\pi)^3}
    e^{-(q/\Lambda)^{2n}} q^{-1}
    (q/\tilde{\Lambda}_\ast)^{-1 + is_0}} \\
    & = -\frac{a_n^{1/2n}}{\Lambda^2}
    \left\{\Re\left[\left(\frac{\Lambda}{\tilde{\Lambda}_\ast}\right)^{is_0}\Gamma\!\left(\frac{is_0 +1}{2n}\right)\right] \right\}^{-1} \\&\times\Re\left[a_n^{-is_0/2n}\left(\frac{\Lambda}{\tilde{\Lambda}_\ast}\right)^{{is_0}}\Gamma\!\left(\frac{is_0 - 1}{2n}\right)\right] \;,
\end{aligned}
\end{equation}
where $a_n \equiv (1 + 2^{-2n})$ and we have used the integral
\begin{align}
    \int_0^{\infty} dx \, e^{-a x^{2n}} x^{b} = \frac{1}{2n} a^{-\frac{b+1}{2n}} \Gamma\!\left(\frac{b+1}{2n}\right) \;.
\end{align}
Under this approximation, $\tilde{\Lambda}_\ast$ is related to  $\Lambda_\ast$ in Eq.~\eqref{H0_0_expr} by
\begin{equation}
    \varphi_0=s_0\ln\tilde{\Lambda}_\ast/\Lambda_\ast \simeq \frac{\pi}{2} -\delta_0 + \arg\!\left(\Gamma\left(\frac{is_0 
    +1}{2n}\right)\right) \;.
\end{equation}
The parameters $\delta_0$ and $h_0$ can be extracted from
Eq.~\eqref{eq:h_general_reg_approx} as 
\begin{align}
    \delta_0 &\simeq \frac{1}{2}\arg\!\left(\left(1 + 2^{-2n}\right)^{is_0/2n}\, \frac{\Gamma\!\left(\frac{is_0 +1}{2n}\right)}{\Gamma\!\left(\frac{is_0 -1}{2n}\right)}\right)\;,
    \label{eq:delta0approx}
    \\
    h_0 &\simeq \left(1 + 2^{-2n}\right)^{1/2n} \, \frac{\left|\Gamma\!\left(\frac{is_0 - 1}{2n}\right)\right|}{\left|\Gamma\!\left(\frac{is_0 +1}{2n}\right)\right|} \;.
    \label{eq:h0approx}
\end{align}
Using the Laurent series for the Gamma function at large $n$, we recover the approximate values for a sharp cutoff in
Ref.~\cite{Bedaque:1998kg}:
\begin{align}
\lim_{n\to\infty} \delta_0 &\simeq \arctan(s_0^{-1}) \;,
\label{appoxdelta0infn}\\
\lim_{n\to\infty}h_0  &\simeq 1 \;,\\
\lim_{n\to\infty}\varphi_0 &\simeq 0 \;.
\end{align}
As $n$ increases, the approximate phase $\delta_0$ converges to the exact sharp-cutoff value, while $\varphi_0$ and $h_0$ slightly deviate from their exact values, 0.05281 and 0.879~\cite{Chen:2025rti}, for a sharp cutoff regulator.

Numerical results for $\delta_0$, $h_0$, and $b_0$, obtained by solving the STM equation \eqref{eq:STM_general_reg} and fitting to Eq. \eqref{H0_0_expr}, are shown as functions of $n$ in Fig.~\ref{fig:params_n}. These results are compared with approximate expressions for $\delta_0$ and $h_0$, as well as with the exact sharp-cutoff values from Ref.~\cite{Chen:2025rti}. The approximate values capture the trend of the numerical results for small $n$, starting about 15\% off and improving progressively. As $n$ increases, the numerical results for $\delta_0$, $h_0$, and $b_0$ converge to values different from those obtained by neglecting the regulators in Eqs.~\eqref{eq:STM_general_reg} and \eqref{eq:Grdef}, and sharply cutting off the integral over $q$ in Eq.~\eqref{eq:STM_general_reg} at $\Lambda$. In contrast, the numerical results are obtained with 
regulators that act on relative momenta (not just $\mathbf{q}$).
The two procedures thus differ in the way they account for momenta comparable to $\Lambda$. While after renormalization observables are insensitive to these details, the running of LECs is not. 
It is, therefore, not surprising that 
the numerical results for super-Gaussian regulators do not converge to the
sharp-cutoff result 
in the large-$n$ limit.

\section{Limit-cycle structure}
\label{sec:structure}

The RG flow we have derived is characterized by the $\beta$-function obtained from Eq. \eqref{H0_0_expr}, 
\begin{align}
    \beta(H_0) &\equiv \Lambda \frac{\partial H_0}{\partial \Lambda} \nonumber\\
    &= \frac{s_0}{h_0 \sin 2 \delta_0} \left(  H_0 - \tilde{H}_{0} \right)\left(  H_0 - \tilde{H}_{0}^\ast \right)\,,
\end{align}
in which $\tilde{H}_{0} \equiv h_0 \exp(2 i \delta_0)$. For $h_0 = 1$ and $\delta_0 = \arctan(s_0^{-1})$, the $\beta$-function reduces to the special form given in
Refs.~\cite{Braaten:2004rn,Pal:2012ioz,Pal:2015wby}. The flow
admits a complex-conjugate pair of fixed points 
$\tilde{H}_{0}$ and $\tilde{H}_{0}^\ast$. Consequently, it does not terminate at a real fixed point and instead acquires a log-periodic character, leading to the limit-cycle behavior and DSI. This is precisely the mechanism underlying the Efimov effect.

The regulator-dependent phase $\delta_0$ controls the separation of the complex pair of fixed points, and the positions (in units of $\Lambda_\ast$) of the poles and zeros of the three-body limit cycle. 
It is known that the poles correspond to the appearance of new deep trimers, which evolve into the Efimov states as $\Lambda$ increases~\cite{Bedaque:1998km}. The zeros are located between  
successive Efimov trimers and mark transitions of 
the effective three-body interaction  
from attractive to repulsive.

\begin{figure}[!tb]
    \centering
    \includegraphics[width=0.95\linewidth]{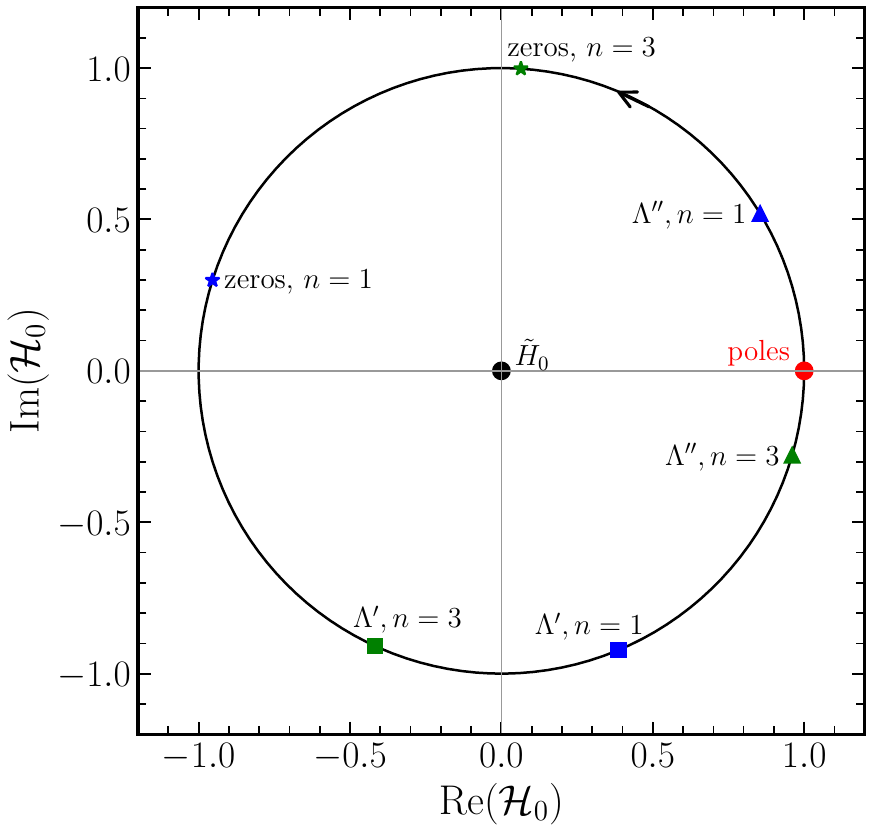}
    \caption{The unit-circle representation of the three-body limit cycle. The arrow indicates the direction of the RG flow as $\Lambda$ increases. The red and black points mark the poles and the fixed point $\tilde{H}_0$ of the limit cycle, respectively. The blue (green) star, square, and triangle on the circle denote the positions of the zeros and of the flow at cutoff values $\Lambda' = 15 \kappa_\ast$ and $\Lambda'' = 800 \kappa_\ast$ on the limit cycle obtained by solving the Faddeev equation with the $n = 1$ ($n = 3$) (super-)Gaussian regulator. }
    \label{fig:unit_circle}
\end{figure}

The RG limit cycle of the three-body coupling is the topological signature of the infinite tower of Efimov states \cite{Horinouchi:2014ata, Horinouchi:2016gqg}. This connection becomes manifest by introducing a complex variable $\mathcal{H}_0$, which is related to $H_0$ through another M\"{o}bius transformation \cite{Pal:2015wby},
\begin{equation}
    \mathcal{H}_0(\Lambda/\Lambda_\ast) \equiv \frac{H_0 - \tilde{H}_0}{H_0 -\tilde{H}_0^\ast} = \exp\left[2i (s_0 \ln (\Lambda/\Lambda_\ast) + \delta_0) \right]\,.
    \label{H0_circle}
\end{equation}
As $\Lambda$ changes, $\mathcal{H}_0$ traces a unit circle in the complex plane, as shown in Fig.~\ref{fig:unit_circle}. The mapping onto a unit circle is not unique. Under the transformation in Eq.~\eqref{H0_circle}, the fixed point $\tilde{H}_0$ is carried to the origin, whereas the other fixed point, $\tilde{H}_0^\ast$, is sent to infinity. If the positions of $\tilde{H}_0$ and $\tilde{H}_0^\ast$ in Eq.~\eqref{H0_circle} are interchanged, the direction of the flow reverses. The poles and zeros of $H_0$ are mapped onto the points (1, 0) and ($\cos 4\delta_0$, $\sin 4\delta_0$), respectively, thus capturing the regulator dependence of the position of zeros relative to poles stemming from $\delta_0$.
Each crossing of the point (1, 0) advances the phase of $\mathcal{H}_0$ by $2\pi$, corresponding to the emergence of a new Efimov trimer. 

The winding number $N_w$, defined as the net number of times $\mathcal{H}_0$ encircles the origin as $\Lambda$ runs from a smaller cutoff value $\Lambda'$ to a larger $\Lambda''$, is
\begin{align}
    N_w = \left\lfloor \frac{s_0}{\pi} \ln \frac{\Lambda''}{\Lambda'} \right\rfloor\,,
\end{align}
in which $\lfloor \cdot \rfloor$ denotes the floor function that rounds down to the nearest integer. The winding number, determined by $s_0$, is regulator independent. 
Since each full winding of $\mathcal{H}_0$ amounts to an additional Efimov trimer, the winding number is equal to the change in the number of Efimov trimers as $\Lambda$ runs from $\Lambda'$ to $\Lambda''$, denoted by $\Delta N_3$, up to a possible offset of 1 if the trajectory does not complete an integer number of full windings,
\begin{align}
    \Delta N_3 &=\left\lfloor\frac{\arg \mathcal{H}_0(\Lambda''/\Lambda_\ast) -\arg \mathcal{H}_0(\Lambda'/\Lambda_\ast)}{2\pi} \right\rfloor \nonumber\\
    &= N_w + \Theta \left[ H_0(\Lambda'/\Lambda_\ast)/h_0 - H_0(\Lambda''/\Lambda_\ast) /h_0 \right]\,,
\end{align}
in which $\Theta[x]$ = 1 if $x>0$ and = 0 if $x\leq 0$. The potential offset depends on $\delta_0$ and $\Lambda_\ast$ (or equivalently $b_0$). An example is provided in Fig.~\ref{fig:unit_circle} for $\Lambda'=15 \kappa_\ast$ and $\Lambda''=800 \kappa_\ast$ on the limit cycles obtained by solving the Faddeev equation with the normal Gaussian ($n=1$) and the sextic super-Gaussian ($n=3$) regulators. For the former, the point (1, 0) lies between $\Lambda'$ and $\Lambda''$ along the direction of the RG flow, yielding an offset of 1. In contrast, in the latter configuration, $\Lambda''$ is located between $\Lambda'$ and (1, 0), and the resulting offset vanishes.

\section{Conclusion}
\label{sec:conclusion}
We have established the universal functional form of the three-body renormalization relation in SREFT for a broad class of separable regulators.  By analyzing the STM equation in the low-energy limit, we demonstrated that the running of the three-body LEC $H_0(\Lambda/\Lambda_\ast)$ universally follows a real M\"{o}bius transformation of $\tan \left(s_0\ln(\Lambda/\Lambda_\ast)\right)$, generalizing the analytic expression previously known only for sharp cutoffs. The generalization is non-trivial and reveals the elegant group structure underlying the three-body limit cycle, which has not been recognized before.

We also showed that the same form applies to the Faddeev equation with separable regulators. This provides the first rigorous derivation for the functional form of the three-body limit cycle within a Hamiltonian framework, a foundation for most few- and many-body approaches, such as the Faddeev–Yakubovsky equations~\cite{Yakubovsky:1967}, variational approaches (for example, the Stochastic Variational Method~\cite{Suzuki:2002stochastic}), and exact-diagonalization methods (for example, the No-Core Shell Model in nuclear physics~\cite{Barrett:2013nh}). Separable regulators are also widely employed in few- and many-body calculations. Our results thus have broad applications and can be used directly as input for more-body studies.  

Our numerical demonstration supports our analysis by validating the universal form across different regulators.  Importantly, while the functional form remains invariant, the regulator dependence of the parameters $\{\delta_0, h_0, b_0\}$ reveals a richer structure in the RG limit cycles than previously recognized. Assuming that the intermediate-momentum form of
the STM wave function remains valid for $k \sim \Lambda$, we derived analytical approximate expressions for these parameters for (super-)Gaussian regulators in the STM equation, extending the analysis of Ref.~\cite{Bedaque:1998kg}. Moreover, by mapping the three-body limit cycle onto a unit circle, we showed that the parameters $\delta_0$ and $b_0$ govern the difference between the increase in the number of Efimov trimers and the corresponding winding number along the RG flow. These findings offer a more complete understanding of three-body renormalization.

The universal form~\eqref{H0_0_expr} is valid in the unitarity limit, where DSI is exact.  When DSI is weakly broken by a finite scattering length $a_0$, $C_0(\Lambda)$ receives additional contributions proportional to powers of $(a_0\Lambda)^{-1}$.  Numerical results show that the ratios between adjacent poles and zeros of $H_0$ gradually approach the universal value $\exp(\pi/s_0) \approx 22.69$ as either the cutoff or the scattering length increases, implying that the universal form we derived holds up to corrections that are suppressed by inverse powers of $a_0\Lambda$, consistent with the analysis in Ref.~\cite{Braaten:2011sz} for a sharp cutoff regulator.  We defer a detailed analysis to future studies.

\begin{acknowledgments}
 LC and PZ thank Ning Sun and Shuyan Zhou for helpful discussions.  FW and SK acknowledge the hospitality of the ECT*, where part of this work was carried out.  LC and PZ are supported by the Shanghai Rising-Star Program under grant number 24QA2700300, the NSFC under grant 12374477, and the Quantum Science and Technology-National Science and Technology Major Project 2024ZD0300101.  XL and SK are supported by the U.S.\ Department of Energy, Office of Science, Office of Nuclear Physics, under Award Number DE-SC0024622, and by the National Science Foundation under Grant PHY-2044632.
 This material is based upon work supported by the U.S.\ Department of Energy, Office of Science, Office of Nuclear Physics, under the FRIB Theory Alliance award DE-SC0013617.
\end{acknowledgments}

\twocolumngrid
\bibliography{main.bbl}

\end{document}